\documentclass[11pt,letterpaper]{article}
\usepackage{systeme} 
\usepackage[utf8]{inputenc}
\usepackage{multirow} 
\usepackage{booktabs} 
\usepackage[english]{babel}
\usepackage{amsmath}
\usepackage{amsfonts}
\usepackage{amssymb}
\usepackage{graphicx}
\usepackage[small,bf]{caption} 
\usepackage[left=3cm,right=3cm,top=2cm,bottom=3cm]{geometry}

\author{Jorge Pinochet}
\title{\textbf{Visualisation of matrix product: Using light to clarify an abstract mathematical concept}}
\begin{document}

\author{Jorge Pinochet$^{1}$, Walter Bussenius Cortada$^{2}$\\ \\
\small{$^{1}$\textit{Departamento de Física, Universidad Metropolitana de Ciencias de la Educación,}}\\
\small{\textit{Av. José Pedro Alessandri 774, Ñuñoa, Santiago, Chile.}}\\
\small{$^{2}$\textit{Facultad de Ciencias de la Educación, Universidad de Talca,}}\\
\small{\textit{Oriente 591, Linares, Maule, Chile.}}\\
\small{e-mail: jorge.pinochet@umce.cl}\\}

\date{}
\maketitle

\begin{center}\rule{0.9\textwidth}{0.1mm} \end{center}
\begin{abstract}
\noindent Teaching the noncommutativity of the product of matrices to high school or college level students is a difficult task when approached from a purely formal perspective. The aim of this paper is to present a simple experimental activity for teaching the noncommutativity of the matrix product, based on the Jones calculus, a mathematical formalism for describing polarised light by means of matrices and vectors. This activity can also be useful to introduce students to the use of matrices in physics, and to illustrate how abstract mathematics can become a powerful tool to help us explain and describe the real world. \\ \\


\begin{center}\rule{0.9\textwidth}{0.1mm} \end{center}
\end{abstract}

\maketitle

\section{Introduction}

Who has never heard their high school physics or mathematics teacher say that "the order of the factors does not alter the product"? This phrase, which expresses the technical concept of commutativity, has far transcended the field of mathematics, and is invoked in the most varied contexts to indicate that the result of a set of actions is independent of the order in which they are performed. However, our professor's phrase has a limited range of applicability, both inside and outside the world of physics and mathematics. Think, for example, of the action of putting a wheel on a vehicle: we first put the wheel on and then the bolts, since if we put the bolts on first, we will not be able to put the wheel on. \\
 
Most people who have graduated from high school and who have not pursued a career in the field of exact sciences are only familiar with operations involving integers and real numbers, where commutativity holds, and are unaware of the existence of mathematical objects that do not satisfy this property. Of these objects, perhaps the most widely used by physicists and engineers are matrices, since they allow us to model and describe a wide variety of phenomena. Those who teach physics or mathematics at the school or university level will have realised that explaining the meaning of the non-commutativity of the product between matrices is an arduous task. In our opinion, the main reason for this is the difficulty of visualising and experiencing non-commutativity, which leads to a purely formal approach that does not contribute to significant learning.\\

The objective of this article is to present a simple experimental activity for teaching the non-commutativity of the matrix product, based on the Jones calculus [1,2], a mathematical formalism that allows us to describe polarized light by means of matrices and vectors. This article may also be useful to introduce students to the use of matrices in physics, and to illustrate how abstract mathematics can become a powerful tool to help us explain and describe the real world.

\section{Theoretical framework: Polarized light and the Jones calculus}

In general, the light we receive from a light bulb or flashlight has an electric field vector representing an electromagnetic wave vibrating in all directions with equal probability. Under these conditions, the light is said to be \textit{unpolarized} [3]. If this light is passed through a linear polarizer, the electric field will be selectively filtered, allowing the plane of vibration to be transmitted in only one direction, while the remaining planes of polarisation will be blocked (see Fig. 1). The polarized light therefore has lower intensity than the original light [4]. Depending on the orientation of the linear polarizer, the plane of polarization can take any spatial orientation, but for our purposes we will consider four directions that we can conventionally designate as vertical ($V$), horizontal ($H$), diagonal at $+45^{o}$ ($D_{+}$), or diagonal at $-45^{o}$ ($D_{-}$, anti-diagonal ), as illustrated in Fig. 2.

\begin{figure}[h]
  \centering
    \includegraphics[width=0.5\textwidth]{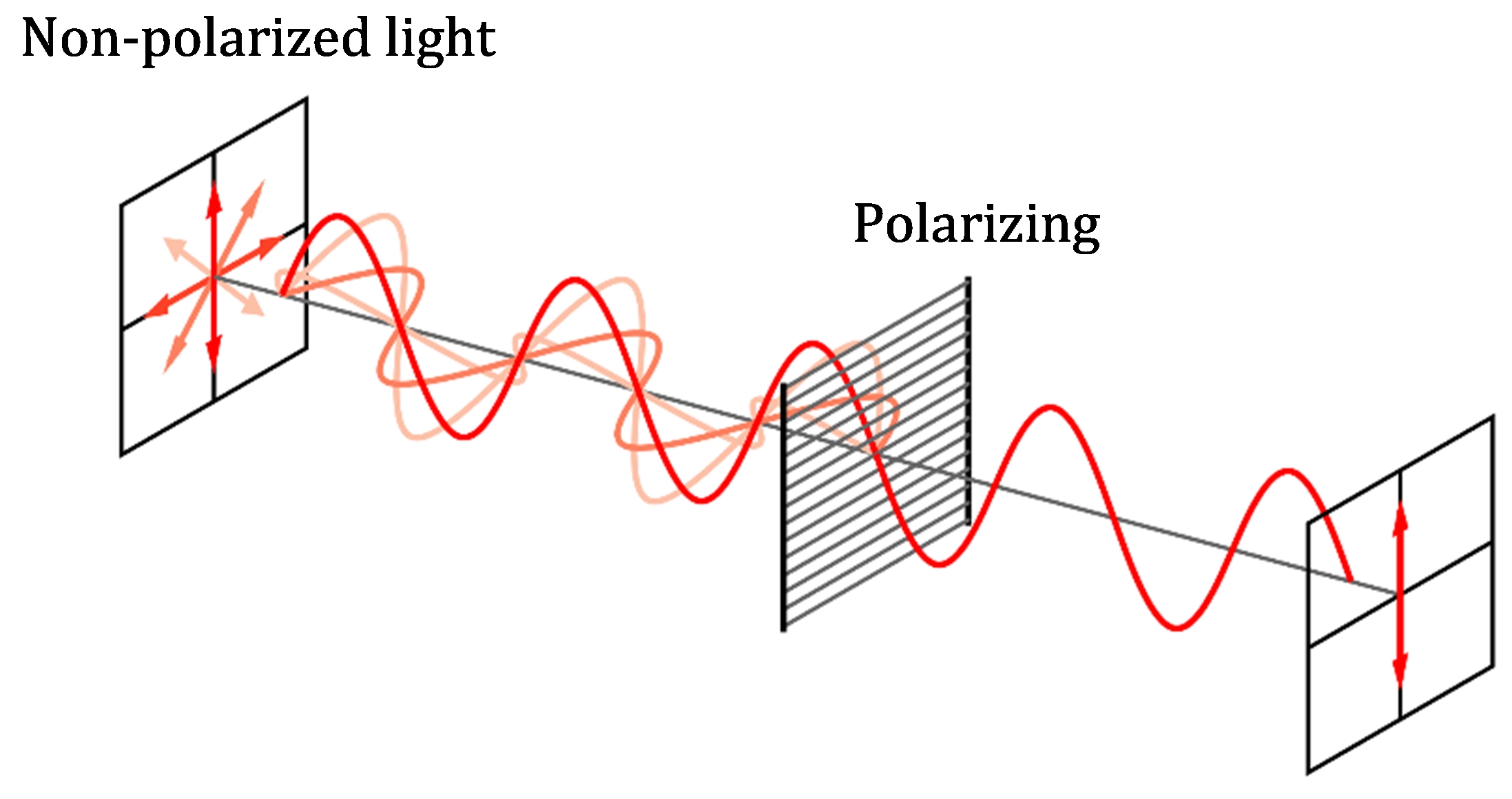}
  \caption{Effect on light of a wire grid linear polarizer (Wikipedia.org). The transmission axis of the grid is perpendicular to the wires due to the presence of electrons in the material, which can only move horizontally, absorbing the electric field in that direction.}
\end{figure}

When vertically polarised light passes through a linear polarizer that makes a certain angle with the vertical, the intensity of the transmitted light will be reduced, since only a component of the original electric field can pass through the polariser. If vertically polarized light passes through a horizontal polarizer, no light will be transmitted, because the vertically vibrating electric field has no horizontal component [3].\\

\begin{figure}[h]
  \centering
    \includegraphics[width=0.2\textwidth]{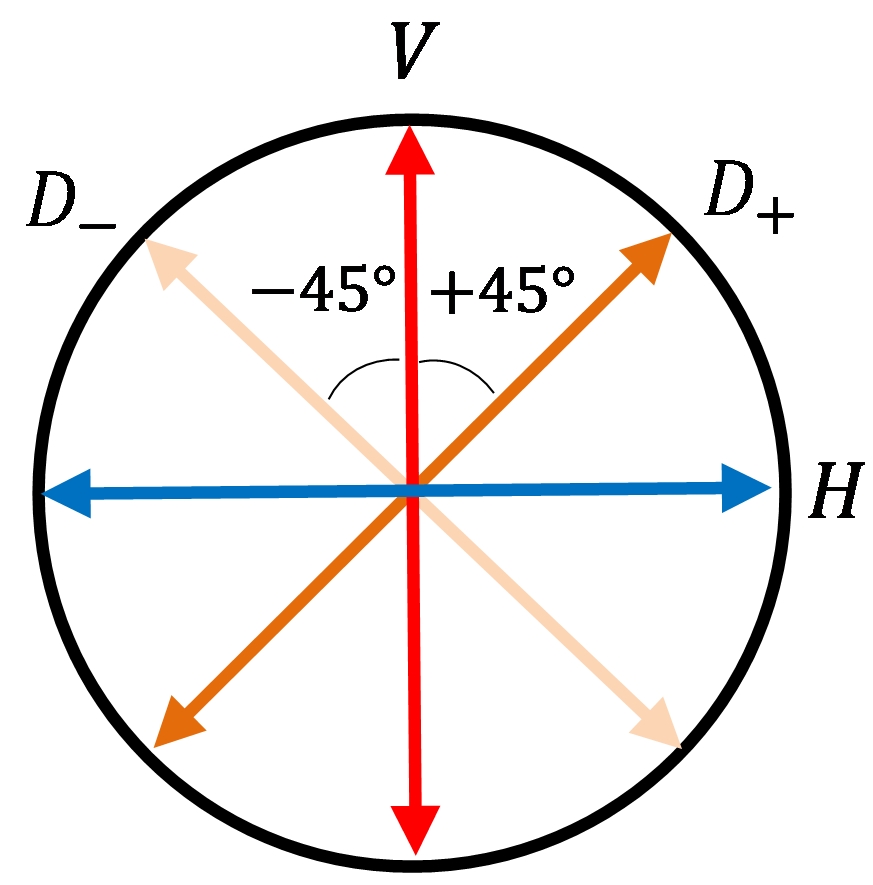}
  \caption{Four possible orientations of a linear polarizer.}
\end{figure}

The Jones calculus is a mathematical formalism devised by the physicist R. Clark Jones, which allows light to be described when it is fully polarized, that is, when the electric field vector only vibrates in one well-defined linear direction [1,2]. Polarized light is represented by two-component vectors called the \textit{Jones vectors}, and the effects of polarizers on light are represented by $2\times 2$ matrices called \textit{Jones matrix}.\\

\begin{table}[htbp]
\begin{center}
\caption{Jones vectors for linear polarization.}
\begin{tabular}{l l l} 
\toprule
\textbf{Polarization} & \textbf{Jones vector} & \textbf{Notation} \\
\midrule
Horizontal linear polarized & $\begin{pmatrix} 1 \\ 0 \\ \end{pmatrix}$ & $\vert H \rangle$ \\ 
\midrule
Vertical linear polarized & $\begin{pmatrix} 0 \\ 1 \\ \end{pmatrix}$ & $\vert V \rangle$ \\ 
\midrule
Diagonal linear polarized ($+45^{o}$) & $\frac{1}{\sqrt{2}} \begin{pmatrix} 1 \\ 1 \\ \end{pmatrix}$ & $\vert D_{+} \rangle$ \\
\midrule
Anti-diagonal linear polarized ($-45^{o}$) & $\frac{1}{\sqrt{2}} \begin{pmatrix} 1 \\ -1 \\ \end{pmatrix}$ & $\vert D_{-} \rangle$ \\
\bottomrule
\end{tabular}
\label{Jones vectors for linear polarization}
\end{center}
\end{table}

Table 1 shows the Jones vectors for linear polarization in each of the four directions defined above ($H, V, D_{+}, D_{-}$). For simplicity, we will not consider the case of circular polarization. The table shows that Jones vectors can be represented in the standard ket notation, which is analogous to that used in quantum mechanics to describe the state vector of a physical system.\\

\begin{figure}[h]
  \centering
    \includegraphics[width=0.5\textwidth]{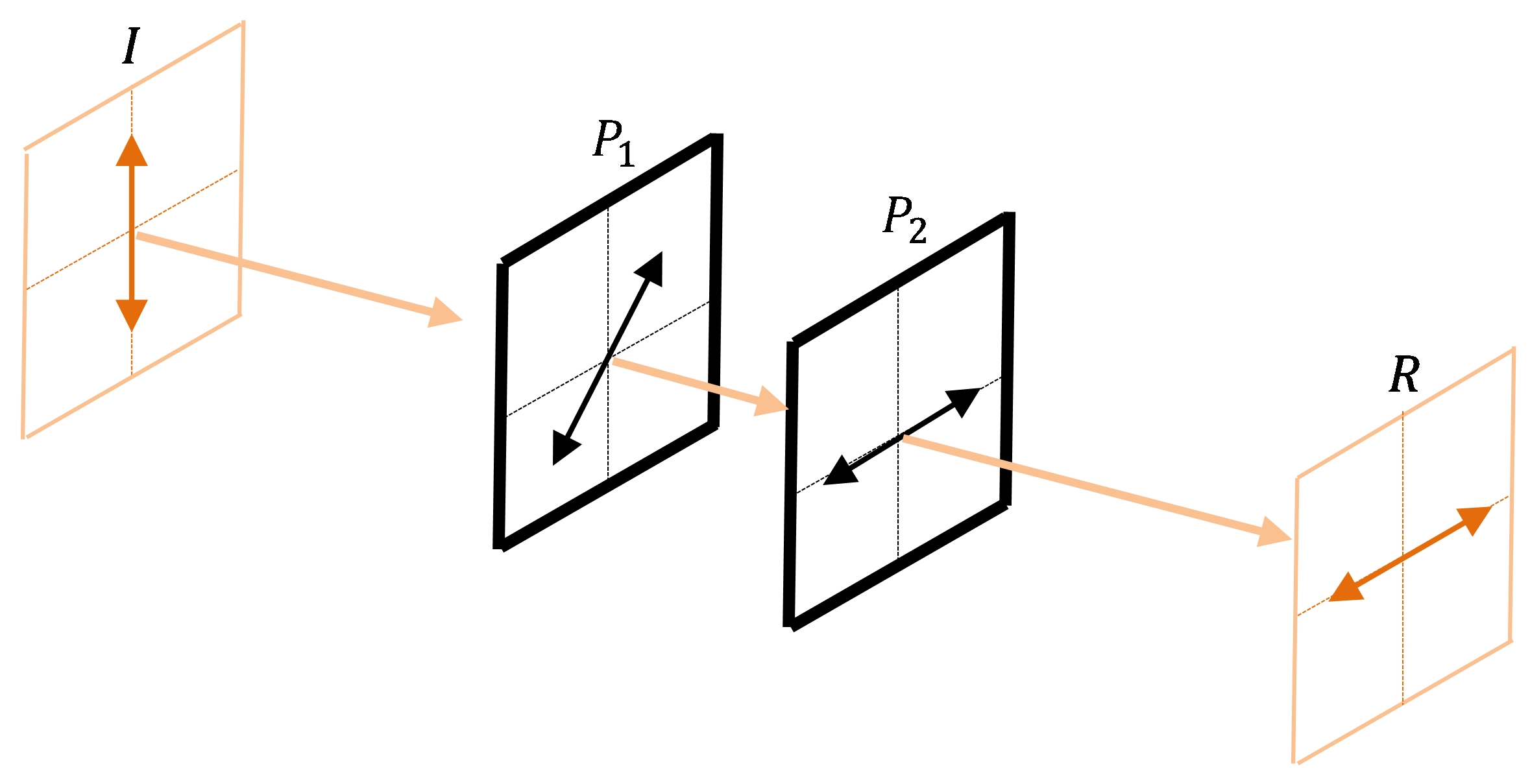}
  \caption{Light with initial polarization $I$ passes through two polarizers $P_{1}$ and $P_{2}$, producing light that is polarized in the direction $R$.}
\end{figure}

Table 2 shows the Jones matrix for the same cases as in Table 1, where again we have omitted circular polarization.\\

\begin{table}[htbp]
\begin{center}
\caption{Jones matrix for linear optical elements.}
\begin{tabular}{l l l} 
\toprule
\textbf{Polarization} & \textbf{Jones matrix} & \textbf{Notation} \\
\midrule
Linear polarizer with horizontal transmission & $\begin{pmatrix} 1 & 0 \\ 0 & 0 \\ \end{pmatrix}$ & $\left[ H \right]$ \\ 
\midrule
Linear polarizer with vertical transmission & $\begin{pmatrix} 0 & 0 \\ 0 & 1 \\ \end{pmatrix}$ & $\left[ V \right]$ \\ 
\midrule
Linear polarizer with diagonal transmission ($+45^{o}$) & $\frac{1}{2} \begin{pmatrix} 1 & 1 \\ 1 & 1 \\ \end{pmatrix}$ & $\left[  D_{+} \right] $ \\
\midrule
Linear polarizer with anti-diagonal transmission  ($-45^{o}$) & $\frac{1}{2} \begin{pmatrix} 1 & -1 \\ -1 & 1 \\ \end{pmatrix}$ & $\left[  D_{-} \right] $ \\
\bottomrule
\end{tabular}
\label{Jones matrices for linear optical elements}
\end{center}
\end{table}

When light passes through a polarizer, the resulting polarization is found by taking the product of the Jones matrix of the polarizer and the Jones vector of the incident light [1,2]. For example, if we have light polarized in an initial direction $I$ with Jones vector $\vert I \rangle$, passing through a polarizer $P_{1}$ represented by the matrix $\left[ P_{1} \right] $, the resulting polarization $\vert R \rangle$ is:

\begin{equation} 
\left[ P_{1}\right] \vert I \rangle = \vert R \rangle.
\end{equation}

If we now assume that the light polarized in the initial direction $I$ first passes through a polarizer $P_{1}$ and then a polarizer $P_{2}$, represented by the matrices $\left[ P_{1 }\right]$   and $\left[ P_{2} \right]$, the resulting polarization $\vert R \rangle$ is obtained as (see Fig. 3):

\begin{equation} 
\left[ P_{2}\right] \left[ P_{1}\right] \vert I \rangle = \vert R \rangle,
\end{equation}

where the product $\left[ P_{1}\right] \vert I \rangle$ is calculated first, and then the product $\left[ P_{2}\right]( \left[ P_{1}\right] \vert I \rangle)$. It is easy to see that the generalisation for $n$ polarizers is:

\begin{equation} 
\left[ P_{n}\right]... \left[ P_{3}\right] \left[ P_{2}\right] \left[ P_{1}\right] \vert I \rangle = \vert R \rangle,
\end{equation}

In the next section, we will see in more detail how the Jones matrix and vectors are used to describe polarized light, and how the empirical observation of the polarization of light allows us to experiment and visualise the non-commutativity of the product between matrices.

\section{The experiment}

The materials required for the experiment are:

-One flashlight.\\
-Three linear polarizers (can be purchased as pieces of glass or mica).\\
-One translucent screen.\\

The experimental setup consists of a flashlight (which produces unpolarised light), a vertical polarizer $V$, a diagonal one $D_{+}$, and a horizontal one $H$ (see Fig. 4). The polarizer $V$ is maintained in a fixed position throughout the experiment, and its only function is to generate a beam of vertically polarized light that provides the input for the experiment. The output comes from the light passing through the polarizers $D_{+}$ and $H$, whose relative position is exchanged during the experiment. If we go back to the sentence “the order of the factors does not alter the product”, we will immediately see that $D_{+}$ and $H$ represent the factors, and the light that emerges from the last polarizer is the product. Hence, the objective of the experiment is to study whether switching or changing the order of the polarizers/factors $D_{+}$ and $H$ generates any alteration in the product/light.\\

We start by analysing the configuration of Fig. 4a. If vertically polarized light passes through the $D_{+}$ polarizer first, only a fraction of the light will pass through, so a reduction in light intensity occurs. If the light subsequently passes through the polarizer $H$, the intensity of the light will again be reduced. In this configuration, the order of the factors/polarizers $D_{+}$ and $H$ results in light emerging, as shown in Fig. 4a.\\

\begin{figure}[h]
  \centering
    \includegraphics[width=0.7\textwidth]{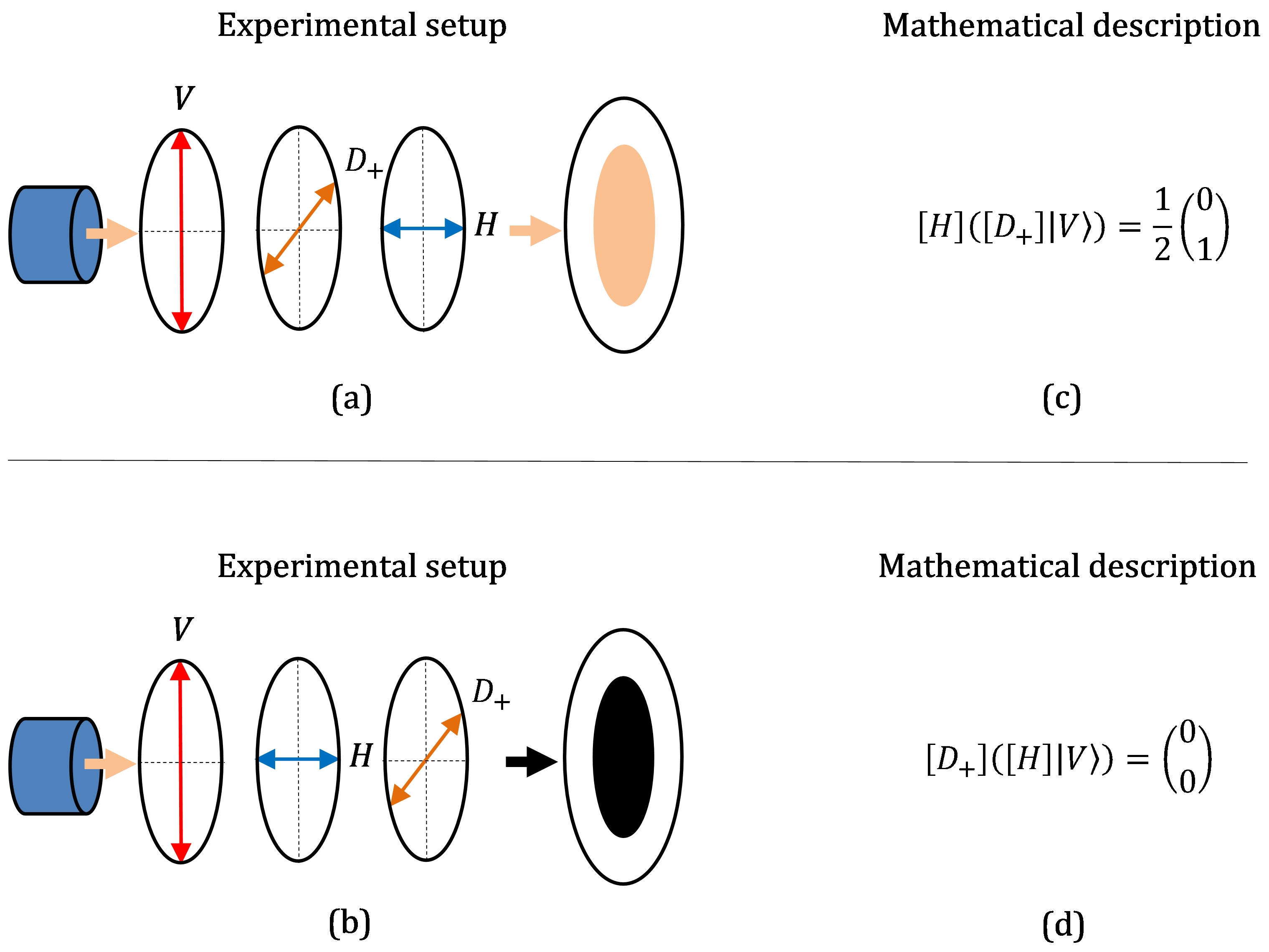}
  \caption{(a) Vertically polarized light from $V$ passes through a diagonal polarizer $D_{+}$ and a horizontal polarizer $H$. The result is that light reaches the screen. (b) Mathematical description when light reaches the screen. (c) Vertically polarized light from $V$ passes through a horizontal polarizer $H$ and a diagonal polarizer $D_{+}$. The result is that no light reaches the screen. (d) Mathematical description when no light reaches the screen.}
\end{figure}

Let us now analyse the configuration in Fig. 4b, and change the order of the factors/polarizers $D_{+}$ and $H$. Under these conditions, the light that is vertically polarized through $V$ is first passed through $H$; since this is a horizontal polariser, it will completely prevent the passage of light, so no light will emerge from $D_{+}$ either. The result will be darkness. Fig. 4b illustrates this situation.\\

\begin{figure}[h]
  \centering
    \includegraphics[width=0.7\textwidth]{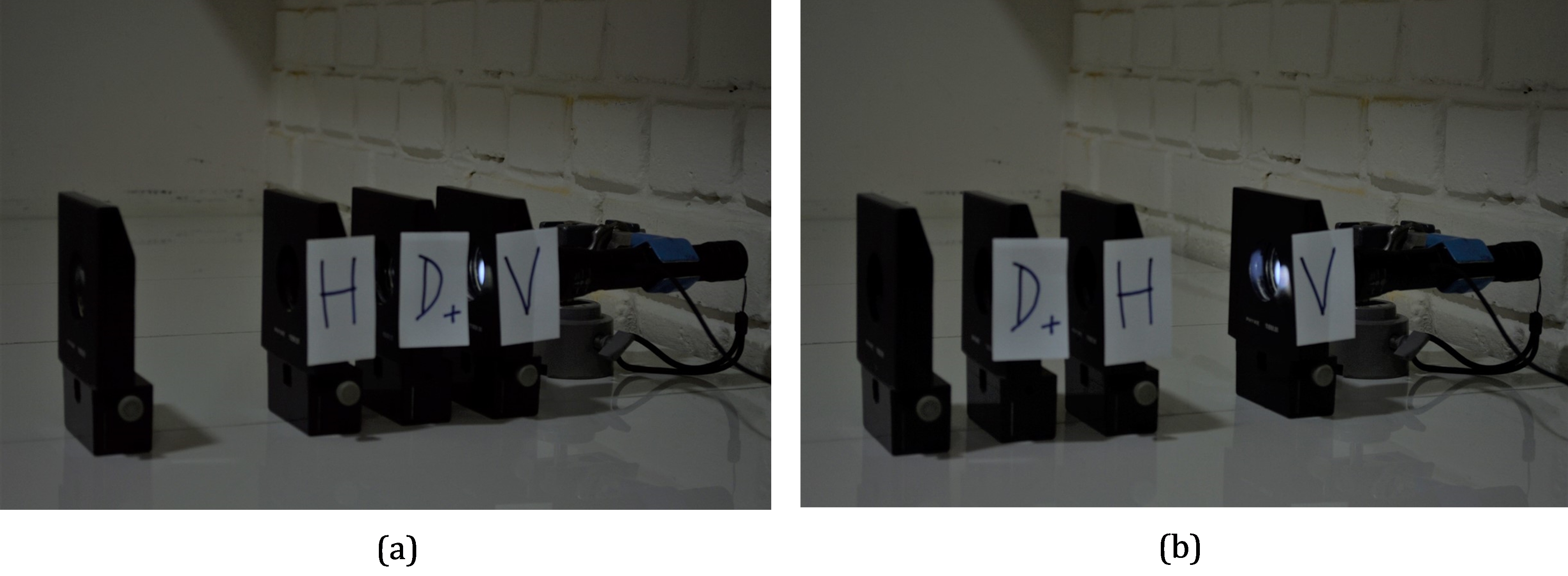}
  \caption{The two experimental configurations of the activity: (a) Configuration illustrated in Fig. 4a. (b) Configuration illustrated in Fig. 4b.}
\end{figure}

Clearly, a change in the order of the factors/polarizers $D_{+}$ and $H$ modifies the product (light in the first case, darkness in the second). This result can also be seen in photographs of the experimental setup (Fig. 5), where the change or commutation of $H$ and $D_{+}$ is shown. It is worth noting that in practice, it is only necessary to move one of the polarizers, which is placed in front or behind the one that remains stationary. Fig. 5a corresponds to the situation illustrated in Fig. 4a, and Fig. 5b corresponds to the situation in Fig. 4b. Fig. 6 clearly reveals the result of switching the polarizers. In Fig. 6a, which corresponds to the configuration in Fig. 4a, light emerges; in Fig. 6b, which corresponds to the configuration in Fig. 4b, no light emerges. The photographs in Fig. 6 were taken facing directly into the light falling on a translucent screen.\\ 

\begin{figure}[h]
  \centering
    \includegraphics[width=0.4\textwidth]{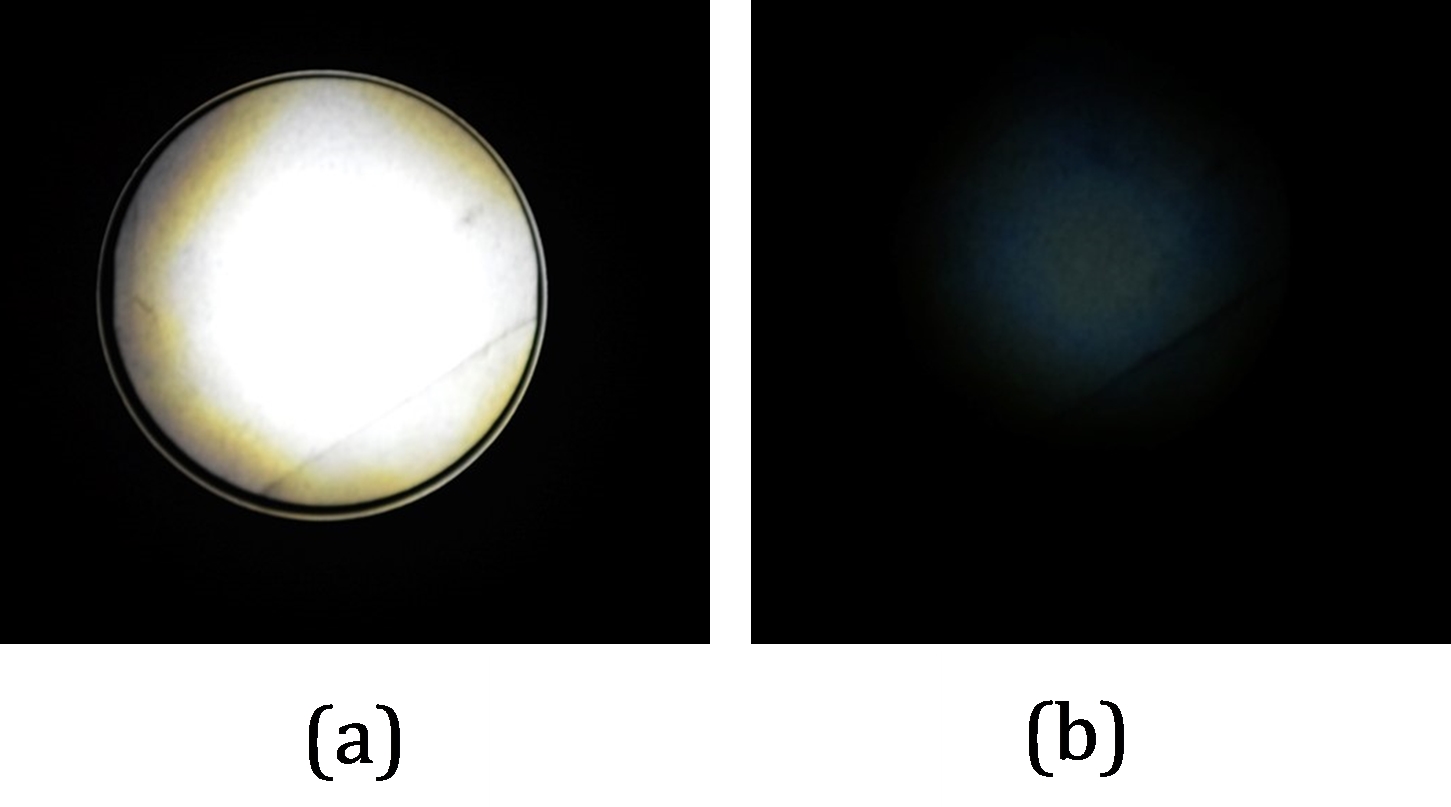}
  \caption{Results of the two experimental activities: (a) In the configuration in Fig. 4a, light reaches the screen. (b) In the configuration in Fig. 4b, no light reaches the screen.}
\end{figure}

Let us now describe mathematically by means of vectors and Jones matrix the results illustrated in the previous figures. According to Table 1, the vector that represents the vertical polarization state of light is:

\begin{equation} 
\vert V \rangle = \begin{pmatrix} 0 \\ 1 \\ \end{pmatrix}.
\end{equation}

According to Table 2, the polarizer $D_{+}$ is represented by the matrix $\left[ D_{+}\right]$ and the horizontal polarizer $H$ is represented by the matrix $\left[ H \right]$, where: 

\begin{equation} 
\left[ D_{+}\right] = \frac{1}{2} \begin{pmatrix} 1 & 1 \\ 1 & 1 \\ \end{pmatrix}, \ \left[ H \right] = \begin{pmatrix} 1 & 0 \\ 0 & 0 \\ \end{pmatrix}.
\end{equation}

We start with the situation illustrated in Fig. 4a, where vertically polarized light passes first through $D_{+}$ and then through $H$. Mathematically, this situation is described as:

\begin{equation} 
\left[ H \right] (\left[ D_{+}\right] \vert V \rangle )= \begin{pmatrix} 1 & 0 \\ 0 & 0 \\ \end{pmatrix} \left[ \frac{1}{2} \begin{pmatrix} 1 & 1 \\ 1 & 1 \\ \end{pmatrix} \begin{pmatrix} 0 \\ 1 \\ \end{pmatrix} \right] = \frac{1}{2} \begin{pmatrix} 1 \\ 0 \\ \end{pmatrix}.
\end{equation}

This result corresponds to horizontally polarized light that emerging attenuated from $H$.\\

Let us now consider the situation illustrated in Fig. 4b, where we change the order of the polarizers $D_{+}$ and $H$. In this configuration, vertically polarized light passes first through $H$ and then through $D_{+}$. Mathematically, this situation is described as:

\begin{equation} 
\left[ D_{+} \right] (\left[ H \right] \vert V \rangle)= \frac{1}{2} \begin{pmatrix} 1 & 1 \\ 1 & 1 \\ \end{pmatrix} \left[ \begin{pmatrix} 1 & 0 \\ 0 & 0 \\ \end{pmatrix} \begin{pmatrix} 0 \\ 1 \\ \end{pmatrix} \right] = \frac{1}{2} \begin{pmatrix} 0 \\ 0 \\ \end{pmatrix} = \begin{pmatrix} 0 \\ 0 \\ \end{pmatrix}.
\end{equation}

We see that the result is a null vector. No light emerges from the polariser $D_{+}$, so we get darkness, which clearly shows that the matrix operations described by Eqs. (6) and (7) are non-commutative; in other words, the order of the factors does alter the product, which we can summarise mathematically as:

\begin{equation} 
\left[ H \right] (\left[ D_{+}\right] \vert V \rangle ) \neq \left[ D_{+} \right] (\left[ H \right] \vert V \rangle).
\end{equation}

Fig. 4 c) and d) illustrates the previous ideas and shows the two experimental configurations and the corresponding mathematical description in each case.\\

Finally, to clear up any doubts, it is worth asking why we only use the diagonal polarizer at $+45^{o}$, and what results would be obtained with a diagonal polarizer at $-45^{o}$. The answer is simple: since the sign of the angle is a mere convention, the results do not depend on whether the orientation of the polarizer with respect to the vertical is at $+45^{o}$ or $-45^{o}$. The reader can easily check this, both from a mathematical point of view (using the Jones calculation with the matrix $\left( D_{-} \right]$), and from an experimental point of view (by performing the activity with the diagonal polariser oriented at $-45^{o}$). In other words, regardless of the orientation of the diagonal polariser, we can verify that the matrix product is non-commutative, and that the order of the factors does alter the product.

\section{Conclusions}

In our opinion, the most important aspects of the activity presented here are firstly that it allows us to experiment with and visualise an abstract mathematical property, and secondly, that it exemplifies the use of matrices in physics in a simple way. In more specific terms, the simplicity of the activity, its low cost, and the short time required to present it in class, which is around five minutes, are noteworthy.\\

In our classroom experience, the proposed experiment has been a valuable tool to promote  better learning of matrix operations, and to demonstrate the way in which an abstract mathematical formalism allows us to model and describe the real world. We therefore believe that this work represents a contribution to the work carried out by physics and mathematics teachers.

\section*{Acknowledgments}
I would like to thank to Daniela Balieiro for their valuable comments in the writing of this paper. 

\section*{References}

[1] G.R. Fowles, Introduction to Modern Optics, 2th ed., Dover Publications, New York, 1989.

\vspace{2mm}

[2] E. Hecht, Optics, 4th ed., Adison Wesley, San Francisco, 2002.

\vspace{2mm}

[3] P.A. Tipler, Physics for Scientists and Engineers, 5th ed., W. H. Freeman and Company, New York, 2004.

\vspace{2mm}

[4] R.A. Serway, J.W. Jewett, Physics for Scientists and Engineers with Modern Physics, 8th ed., Thomson, Cengage Learning, 2010.

\end{document}